\documentstyle[12pt,aasms4]{article}

\begin{document}

\def\etal{{\it et al.\/\ }}
\def\today{\ifcase\month
 &  &  &  &  &  \or January &  \or February \or March &  & \or April 
 &  &  &  &  &  \or May &  &  &  \or June &  &  \or July &  &  \or August
 &  &  &  &  &  \or September \or October & \or November \or December\fi
 &  &  &  &  &  \space\number\day, \number\year}
\def\ga{\lower 2pt \hbox{$\, \buildrel {\scriptstyle >}\over{\scriptstyle \sim}\,$}}
\def\la{\lower 2pt \hbox{$\, \buildrel {\scriptstyle <}\over{\scriptstyle \sim}\,$}}

\title{\bf THE WEAKNESS OF C IV ABSORBER CLUSTERING
IN KECK HIRES SPECTRA OF ADJACENT QSO SIGHTLINES$^1$}

\author{Arlin P. S. Crotts}

\affil{Department of Astronomy, Columbia University, 538 W.~120th St., 
New York, NY~~10027}

\author{Scott Burles \& David Tytler}

\affil{Department of Physics, and Center for Astrophysics and Space Sciences,
University of California, San Diego, MS~0424, La Jolla, CA~~92093-0424}

\affil{\bigskip
$^1$ Based on observations made at the W.M. Keck Observatory which is operated
as a scientific partnership between the California Institute of Technology and
the University of California; it is made possible by the generous support of
the W.M. Keck Foundation.}

\authoremail{arlin@astro.columbia.edu, scott@cass.ucsd.edu,
tytler@cass.ucsd.edu}

\begin{abstract}

We observe with Keck/HIRES the $z\approx2.5$ QSO triplet 1623+27 in order to
explore on the scale of a Megaparsec the spatial clustering of C~IV absorbers
between adjacent sightlines.
We find this signal to be significantly weaker than the clustering in velocity
on corresponding scales along single sightlines, assuming that the relative
velocity of absorbers is dominated by the Hubble flow.
This indicates that small-scale clustering (200~km~s$^{-1} < \Delta v <
600$~km~s$^{-1}$) of the C~IV absorbers cannot be interpreted in terms of the
positions of the absorbers in space, but must be considered as internal motions
within individual absorbers, or within clusters of absorbers whose internal
velocities dominate over Hubble expansion across the cluster scale.
If the single-sightline signal is due to spatial clustering, it is caused by
absorber clusters smaller than would be implied by their velocities if a Hubble
flow is assumed.
The spatial clustering of C~IV absorbers at $z\approx2$ is consistent with
data on Ly $\alpha$ forest clustering measured in the same way at the same
redshifts.
However, present-day galaxy clustering, evolved back to $z\approx2$, is
consistent with C~IV spatial clustering but perhaps not with that of the
Ly $\alpha$ forest.
Even so, one cannot as yet distinguish the two absorber populations on the
basis of spatial clustering on these small scales.

\end{abstract}


\clearpage

\section{INTRODUCTION}

One of the hallmark distinctions between QSO absorption systems containing
strong C~IV lines and those which do not has been, since early in the history
of their study, the difference in clustering between the two populations as
seen along single lines of sight.
The Ly~$\alpha$ forest has always been shown to be weakly clustered compared to
C~IV absorbers (Webb 1987, Hu et al.~1995, Chernomordik 1995, Cristiani et
al.~1995), and perhaps not detectably clustered at all (Sargent et al.~1980,
Rauch et al.~1992, Lu et al.~1996, Kirkman \& Tytler 1997).
In contrast, even early studies showed that C~IV absorbers cluster significant
in velocity along single sightlines (Young et al.~1982), with perhaps the
best evidence coming from a large compilation of QSO spectra at approximately
100~km~s$^{-1}$ resolution (Sargent et al.~1988, hereafter SSB).
This is usually described by the two-point correlation function $\xi$ which
equals the excess number of absorbers over random expectation found at a
certain location with respect to a given absorber, usually quantified as a
spatial or velocity separation between the two locations (and with $\xi$
normalized by dividing by the random expectation).
Depending on the sample selected, for line-of-sight velocity differences
$\Delta v$ of 200-600~km~s$^{-1}$, values of $\xi$ for C~IV absorbers were
found with $\xi \approx 10$ or larger.

Similar behavior is found in large C IV samples at spectral resolution higher
than that of SSB e.g.~within the largest clustering sample, 10 sightlines at
18 to 40 km s$^{-1}$ FWHM resolution, analyzed by Petitjean and Bergeron
(1994 - PB).
(Other recent works at even higher resolution are based on even fewer
sightlines: Songaila \& Cowie 1996, Fern\'andez-Soto et al.~1996, Rauch et
al.~1996.)
PB, like SSB, find an average $\xi \approx 10$ for $\Delta v
=200$-600~km~s$^{-1}$, dominated by a broad, slowly declining component.
(Specifically, they fit $\xi$ with two components, of widths $\sigma_v=109$ and
525~km~s$^{-1}$, with the broader component containing 71\% of the pair count
over 30-1000~km~s$^{-1}$ and 93\% over 200-600~km~s$^{-1}$.)~
In contrast, even in those papers which found some clustering in the
Ly~$\alpha$ forest, also on approximately these $\Delta v$ scales or slightly
smaller, the signal rarely exceeded $\xi \approx 1$.
This is seen as clear evidence for a difference between these population,
possibly implying distinct origins for the two.

There are caveats to this interpretation which suggest caution in comparing
the single-sightline $\xi$ values of the Ly~$\alpha$ forest and C~IV absorber
populations.
First, it is possible that single-sightline velocity splittings might arise
from internal motions within absorbers, in which case the differences between
the two populations' single-sightline $\xi$ functions are not clearly tied to
their spatial clustering behavior.
SSB argue that, for C~IV absorption arising in galaxy haloes, these velocity
splittings cannot be due to gravitational orbits within these haloes, and
clustering still contributes the dominant portion of the observed $\xi$ on
scales larger than $\Delta v = 200$~km~s$^{-1}$.
Indeed, many papers have modeled QSO absorption-line clustering in terms of
spatial separations indicated by their relative velocities, whereas for highly
over-dense structures, large differences between velocity clustering and
spatial clustering become apparent e.g.~Kaiser (1997).
Models have been proposed, however, where non-gravitational acceleration might
lead to splittings with large $\Delta v$ (York et al.~1986).

Second, the intrinsic width of Ly~$\alpha$ lines, up to $b \approx
60$~km~s$^{-1}$, is much higher than for C~IV because of thermal broadening,
and significant on the scale of the clustering in $\Delta v$ being discussed.
Line-of-sight blending of Ly~$\alpha$ lines does seem to obscure some of the
clustering power seen in their corresponding C~IV lines Fern\'andez-Soto et
al.~(1996).

One way to assess the importance of such effects is to study the clustering of
absorbers in adjacent sightlines close enough together so that the angular
separation between them is less than or comparable to the velocity scales where
clustering is seen or sought in single sightlines,
here assuming that a Hubble expansion law at high redshift can be used to
relate $\Delta v$ and transverse separation.
This addresses all of the above concerns.
First, purely internal velocity splittings cannot shift absorbers to a
different sightline, and, second, blending cannot eliminate all of the small
$\Delta v$ absorber pairs that would otherwise exist between sightlines.
Even if blending decreases or splittings increase the number of close absorbers
pairs between sightlines, there is much less effect on $\xi$ because these
effects also change the total numbers of pairs used to normalize $\xi$.
If all absorbers in a population are equally likely to cluster i.e.~if all
within a population cluster in a way described purely by the same $\xi$, the
effects of blending or line splitting on close pairs and distant pairs cancel.

The 1623+27 QSO triplet discovered by Sramek and Weedman (1978) and 
has been used to measure the spatial two-point correlation function 
(here also denoted by $\xi$) of Ly $\alpha$ absorbers (Crotts 1989, Crotts \&
Fang 1996, with some members of the triplet also observed by Sargent et
al.~1982, and SSB).
We have obtained Keck HIRES spectra of these three QSOs, yielding a sample of
C~IV absorbers large enough and unambiguous enough that a useful comparison
of spatial C~IV clustering can be made to Ly $\alpha$ clustering and
single-sightline C~IV clustering.
These C~IV clustering results are rather different from prior results from
single sightlines alone, which changes our understanding of clustering at high
redshift.

\section{Observations and Analysis}

On the night of 20 May 1996, we used the HIRES spectrograph (Vogt 1994) on
the Keck-1 10m telescope to obtain spectra of the QSO triplet Q1623+27.
Each of the QSOs was observed with the same setup, providing
wavelength coverage from 3872 to 6299\AA.
The observations were performed sequentially over a four hour period, and the
spectrograph was not moved between observations.
We exposed for 5400s on both Q1623.7+268A (KP 76, $V= 18.4$, $z_{em}=2.467$)
and Q1623.9+268 (KP 78, $V=19.4$, $z_{em}=2.607$), and 3000s on Q1623.7+268B
(KP 77, $V=17.0$, $z_{em}=2.521$).
The exposures were taken with a 1.14$^{\prime\prime}\times 7.0^{\prime\prime}$
slit, which gave a resolution of 8~km~s$^{-1}$ and adequate sky coverage.
The images were processed and the spectra were optimally extracted using an
automated routine specifically designed by T.~A.~Barlow for HIRES spectra.
The routine performs baseline subtraction, bias and flat-field corrections, and
utilizes a bright standard star to trace the echelle orders and define the
apertures for extraction.
Thorium-Argon lamp images were obtained immediately after the observations to
provide wavelength calibrations in each echelle order.
The root-mean-square residuals in the wavelength calibration for each echelle
order was less than 0.3~km~s$^{-1}$.
All wavelengths are vacuum values in the heliocentric frame.
Each echelle order was continuum fit with a legendre polynomial
to normalize the unabsorbed QSO flux level to unity.

As an example of these data, we present Figure 1, which shows a particularly
complex C~IV doublet from the faintest QSO, KP 78, fit by 10 components.
The positions of components and best fit flux from VPFIT (Carswell et al.~1992)
were determined; seen in Figure 1 for the $z\approx2.24$ system, along with the
continuum fit.
The spectrum does not have useful SNR in the blue, where the correcponding
Lya lines lie.

The redshifts of C~IV doublets found in these data are listed in Table 1.
We include only those redward of the Ly $\alpha$ forest, and list them according
to the 200~km~s$^{-1}$ ``blended'' sample treated below.
In comparison, the Crotts and Fang (1997) $\Delta \lambda \approx 1.5$\AA\ KPNO
4m sample contains within this redshift interval the two stronger C~IV doublets
in KP 76, all of the KP 77 sample (with 1.878027 and 1.880660 blended together),
and all of the KP 78 sample except 2.115063, with 2.061465 listed as uncertain.
Their uncertain system at $z = 2.40602$ KP 77 appears to be a misinterpretation
of the confused region of C IV and Mg II doublets near 5280\AA.

\section{Results}

%all systems: 20.1 per 500 km/s over first 20000 km/s.
%Linear fit to first 20000 km/s: 18.95 +- 6.71, Linear fit to all: 18.45 +-1.59
%36 in v < 500 km/s, 79 in v < 1000 km/s.  NOT POISSON

%blended (< 200 km/s): 2.1 per 500 km/s over first 20000 km/s.
%Linear fit to first 20000 km/s: 2.37 +- 0.57, Linear fit to all: 2.10 +- 0.15
%4 in v < 500 km/s, 7 in v < 1000 km/s, 13 in v < 2000 km/s.  POISSON?

%A2: 1.375 per 500 km/s over first 20000 km/s:
%Linear fit to first 20000 km/s: 1.69 +- 0.49, Linear fit to all: 1.23 +- 0.12
%1 in v < 500 km/s, 4 in v < 1000 km/s, 9 in v < 2000 km/s.  POISSON?
% SSB: xi (200-600 km/s) = 5.7 +- 0.6

%A4: 0.8 per 500 km/s over first 20000 km/s:
%Linear fit to first 20000 km/s: 1.07 +- 0.32, Linear fit to all: 0.62 +- 0.07
%0 in v < 500 km/s, 2 in v < 1000 km/s, 5 in v < 2000 km/s.  POISSON?
% SSB: xi (200-600 km/s) = 11.5 +- 1.3

Sightline cross-correlating pairs for all C~IV systems results in 36 in the
first bin ($\Delta v < 500$~km~s$^{-1}$), compared to the random expectation
from a linear fit over the first 20000~km~s$^{-1}$ of $18.95 \pm 6.71$ for the
first 500~km~s$^{-1}$ bin.
These counts are obviously highly non-Poisson, so we do not assign an error
estimate to the resulting two point function of $\xi(\Delta v < 500$~km
s$^{-1}) = 0.90$.

It is more reasonable to consider merging C~IV redshifts close to each other
in the same sightline, since it seems likely that these are multiple
representatives of the same absorber.
If we over-correct for this effect, we do not damage the cross-sightline $\xi$,
since blending together systems does not reduce the fraction of pairs between
sightlines due to close cross-sightline $\Delta v$ values, compared to the
total number of cross pairs.
We choose to blend together all systems in the same sightline within
200~km~s$^{-1}$ of each other, starting with the smallest splitting first.
This is the same criterion adopted by SSB, so it leads to a direct comparison.
The cross correlation of this sample (and samples defined by further
criteria) are shown in Figure 2.

In correspondence with SSB, we reduce the sample to only those systems which
would likely have been detected by their survey.
This is a heterogeneous selection in terms of rest equivalent width $W_o$, and
corresponds roughly to $W_o = 0.1$\AA\ for their ``Sample A2'' (which also
excludes all absorbers within 5000~km~s$^{-1}$ of the emission-line redshift.
A homogeneous sample in $W_o$ requires a cut at 0.15\AA\ (their sample ``A4,''
also with $\beta c > $5000~km~s$^{-1}$).

The randomly expected number of pairs in the first 500~km~s$^{-1}$ bin for
each of these subsamples (``Blended,'' ``A2,'' and ``A4'') are $2.37 \pm 0.57$,
$1.69 \pm 0.49$, and $1.07 \pm 0.32$, respectively, whereas the actually
observed number of pairs in the first bin for each sample is 4, 1 and 0,
respectively, leading to $\xi$ values of $0.68^{+1.34}_{-0.52}$,
$-0.41^{+1.39}_{-0.57}$, and $-1^{+1.75}_{-0}$, respectively.
(These are 68\% confidence intervals, corresponding to $\pm 1 \sigma$, assuming
Poisson errors in pair counts, which is close to correct.)

\section{Discussion and Conclusions}

%z=2.15, q0 = 0.5: H0=559.1 km/s, 127" <-> 512 kpc/h
%q0 = 0.1: H0=376.7 km/s, 127" <-> 721 kpc/h

At $z=2.15$ (and for $q_o = 1/2$) the transverse separations between the three
QSOs correspond to velocities in the Hubble flow of 286 to 399~km~s$^{-1}$.
(For $q_o = 0.1$ they are 0.95 times smaller.)~
Therefore, structure dominated by the Hubble flow over 200 to 600~km~s$^{-1}$
would contribute to clustering on these scales, and correspond to proper
separations of 0.36 to 1.07~$h^{-1}$~Mpc.
Since separations between the sightlines are smaller than this (0.51 to
0.71~$h^{-1}$~Mpc), one must add a perpendicular (line of sight) velocity
component up to about 500~km~s$^{-1}$ (although more typically about
200~km~s$^{-1}$).
Correlational activity from such a signal should be restricted to the first
500~km~s$^{-1}$ bin in Figure 2.

Nevertheless, the single sightlines over 200-600~km~s$^{-1}$ and the multiple
sightlines for $v<500$~km~s$^{-1}$ probe slightly different volumes around each
absorber.
We can evaluate the importance of the different sampling regions by considering
the analytic fit by PB to the number of pairs in these velocity intervals.
They find that the number of pairs is well-approximated by the sum of two
gaussians, with standard-deviation velocity widths of 109 and 525 km s$^{-1}$,
and with the wider gaussian contributing 30\% of the number of pairs to the sum
of the gaussians at their peak at zero velocity.
When we integrate this distribution over the SSB sampling volume covering
200-600 km s$^{-1}$, we find 1.15 times as many pairs as when we integrate over
the triple-sightline sampling volume.
This is a relatively minor effect which we neglect hereafter, but one which
tends to lower slightly the discrepancy we discuss below.

The actual value seen by SSB for the A2 sample is $\xi (200-600$~km~s$^{-1}) =
5.7 \pm 0.6$, which should be compared to our $\xi=-0.41^{+1.39}_{-0.57}$,
and for A4 $\xi (200-600$~km~s$^{-1}) = 11.5 \pm 1.3$, which is even more
directly comparable to our $\xi=-1^{+1.75}_{-0}$.
The A2 result is inconsistent at about the $4\sigma$ level, while the A4 result
is discrepant by about $6 \sigma$, both in the sense that the absorbers are
less clustered in adjacent sightlines than is predicted by the
single-sightlines result assuming pure Hubble flow.

%ALL: 131 (0-600) vs 101 (600-10200), BLEND: 3 vs 20, A2: 2 vs 14, A4:2 vs 3
We confirmed that the line-of-sight clustering in our three spectra are
consistent with those in the SSB sample of 55 QSOs.
We constructed the C~IV redshift auto-correlation function for $\Delta v <
600~km~s^{-1}$ along each of the three sightlines taken individually, then
summed together.
For all systems, one finds $\xi_{auto} = 19.8 \pm 2.4, ~ 1.4^{+2.3}_{-1.3}
~ 1.3^{+3.0}_{-1.5}$, and $9.7^{+17.5}_{-9.0}$, for all C~IV systems, blended
systems, ``A2'' and A4 samples, respectively, in the first 600~km~s$^{-1}$ bin.
These are constructed using the average number of pairs in 600~km~s$^{-1}$ bins
with 600~km~s$^{-1} < \Delta v < 10200~$km~s$^{-1}$.
These measurements are consistent with their corresponding SSB values, albeit
at much poorer $S/N$ due to the smaller number of QSO sightlines.

The inconsistency of the two-point correlation function derived from Figure 2
with that of SSB implies that single sightline correlation functions cannot be
used to study the {\it spatial} clustering of absorbers on velocity scales of
several hundred km~s$^{-1}$.
This may be due either to internal velocities within absorbers that are caused
by non-gravitational processes, or by motion within gravitational potentials
that have separated from the Hubble flow.
These structures, either individual absorbers or clusters of absorbers, must
be small enough to add little clustering power on scales of 0.5 to 1.1 $h^{-1}$
Mpc.
In either case, most of the line-of-sight correlational power is due to
behavior not described by the absorber positions alone, but some peculiar
motion.
Line-of-sight absorber correlation functions should not be compared directly to
galaxy correlation functions usually expressed as $\xi (r)$ in terms of a
radial separation vector $r$ in space.

This spatial clustering of C~IV absorbers is much weaker than would be expected
for galaxies at $z=0$.
A direct comparison involves averaging the galaxy correlation function $\xi =
(r/r_o)^{-\gamma}$, where we assume $\gamma = 1.8$ and $r_o = 7~h^{-1}$~Mpc
(derived from Park et al.~1994, although there are lower $r_o$ values for
different samples e.g.~Fisher et al.~1994).
This is averaged over a line segment extending from the tangent point at
closest approach (0.512, 0.593, and 0.714 $h^{-1}$ Mpc for each of the
sightline pairs) and extending to the point at $\Delta v = 500$~km~s$^{-1}$
(1.031, 1.073, and 1.144 $h^{-1}$ Mpc, respectively).
Averaged over all three sightlines, $\bar{\xi} = 57.4$.
Assumably, this can be back-evolved to $z=2.15$ with stable hierarchical
clustering formalism (Davis \& Peebles 1977) if $\bar{\xi} >> 1$, according to
$\bar{\xi} (z) = \bar{\xi}(0) (1+z)^{-3} = 1.84$.
(Formally, this assumes $q_o = 1/2$, but remaining non-linear over all
relevant $z$, is a close approximation for other cosmilogical densities.)~
Note that high-$z$ clustering (Hudon \& Lilly 1996) measured at $z=0.48$
corresponds to $3.2 < \bar{\xi} < 7.7$, and extrapolates to $0.33 < \bar{\xi} <
0.80$ at $z=2.15$ assuming stable clustering.
These result is consistent with any of the comparable values obtained above for
C~IV absorbers.

Even though there are large differences between the line-of-sight clustering of
the Ly $\alpha$ forest and C~IV systems, their clustering power between
different lines of sight is more similar.
The strength of C~IV clustering is consistent with that of the Ly $\alpha$
forest at similar redshifts.
Crotts \& Fang (1996) show, for these same sightlines at nearly the same
redshift $\langle z \rangle = 2.14$, that $\xi = 0.86 \pm 0.35$ for
Ly $\alpha $ lines with $W_o = 0.4$\AA\ and $\Delta v < 200$~km~s$^{-1}$.
For $\Delta v < 500$~km~s$^{-1}$, there are 51 pairs observed versus 19
expected, implying $\xi = 0.31\pm0.18$ ($1\sigma$).
Measured in this way, it is unclear that C~IV clustering is stronger than
Ly $\alpha$ clustering.
However, Ly $\alpha$ spatial clustering is less than the expectation for
galaxies assuming stable hierarchical clustering ($\bar{\xi} = 1.61$), but not
necessarily weaker than when we start from the Hudon \& Lilly result.

One remaining question is whether the structure we probe on 0.5 Mpc scales
might actually probe the scale of individual absorbers.
We are fairly confident that this analysis of $\xi$ addresses more the
clustering of absorbers than some measure of their characteristic size.
The size of C~IV absorbers is indicated by the transverse separation at which
absorbers in one sightline have high probability of appearing in the adjacent
sightline.
For gravitationally-lensed QSOs (Steidel \& Sargent 1991) and for distinct QSO
pairs (Crotts et al.~1994), strong correspondence between adjacent absorption
lines indicates C~IV absorber sizes of only a few tens of kiloparsecs.
On scales smaller than this, absorbers are presumed to merge.
On scales larger than this, up to the 0.5~$h^{-1}$~Mpc sampled by the QSO
triplet, it is still possible that motion within objects that have collapsed
out of the Hubble flow might still be responsible for much of the
clustering signal for 200~km~s$^{-1} < \Delta v < 600$~km~s$^{-1}$ reported by
SSB.
Calculating $\bar{\xi}$ for the galaxy two-point correlation function at
$z=2.15$, assuming stable hierarchical clustering development, one finds
$\bar{\xi}=33$ for separations (along a sightline) of 40~$h^{-1}$~kpc to
0.5~$h^{-1}$~Mpc, still larger than $\xi$ found for the A2 or A4 samples of
SSB.
The SSB 200-600~km~s$^{-1}$ clustering signal might plausibly be explained as
clusters of absorbers smaller than 0.5~$h^{-1}$~Mpc with internal velocities of
a few hundred km~s$^{-1}$.
Indeed, high resolution simulations of fragments collapsing ultimately into
galaxies at $z \approx 0$ show that these fragments at $z \approx 3$ subtend
such spatial scales e.g.~Rauch, Haehnelt \& Steinmetz (1997).

These conclusions are based on a single group of sightlines, and such close
groupings of reasonably bright, sufficiently high $z$ QSOs are extremely rare.
Nonetheless, a larger sample to check and refine these conclusions is desired.

\acknowledgments

This research was supported in part by a David and Lucile Packard Foundation
Fellowship to A.C. and NSF grant AST-9420443 and NASA grant NAGW-4497 to D.T.~
We thank P.J.E.~Peebles and C.~Cress for helpful discussion.
We thank Tom Barlow and Bob Carswell for software which made the data reduction
and analysis possible.

\clearpage

\newpage 

\noindent Figure 1 shows a complex $z\approx2.24$, C~IV doublet from KP 78, fit
by 10 Voigt components.

\bigskip
\bigskip

\noindent Figure 2 shows the cross-correlation between pairs of the three
sightlines, for three restricted samples:
all systems after blending within 200~km~s$^{-1}$ (dashed curve),
blended systems with rest equivalent width $W_o > 0.10$ for C~IV $\lambda$1548,
in close analogy to SSB sample A2 (horizontally-striped bars), and
blended systems with rest equivalent width $W_o > 0.15$, as in SSB sample A4.

\newpage

\begin{deluxetable}{cccl}
\tablecaption{C~IV ABSORPTION SYSTEMS FOUND IN QSO TRIPLET 1623+27}
\tablewidth{0pt}
\tablehead{
\colhead{QSO} &
\colhead{Blended} &
\colhead{C~IV~$\lambda$1548 Rest} &
\colhead{Component Redshifts} \\
\colhead{} &
\colhead{System z} &
\colhead{EW, $W_o$ (\AA)} &
\colhead{}
} 
\startdata
{\bf KP 76} & 1.845177 & $0.055\pm0.006$ & 1.845177\nl
 & 2.112378 & $0.313\pm0.007$ & 2.111746, 2.112022, 2.112872\nl
 & 2.156867 & $0.071\pm0.008$ & 2.156484, 2.157249\nl
 & 2.245817 & $0.178\pm0.008$ & 2.245372, 2.246084, 2.246438\nl
% & & \nl
{\bf KP 77} & 1.878027 & $0.070\pm0.003$ & 1.878027\nl
 & 1.880660 & $0.144\pm0.003$ & 1.880084, 1.881235\nl
 & 1.972602 & $0.124\pm0.004$ & 1.972276, 1.972929\nl
 & 2.050746 & $0.662\pm0.006$ & 2.049659, 2.049868, 2.050201, 2.051020, 2.051807, 2.052194\nl
 & 2.052938 & $0.487\pm0.006$ & 2.052644, 2.052866, 2.053120\nl
 & 2.161619 & $0.307\pm0.005$ & 2.161104, 2.161317, 2.161332$^a$, 2.162024\nl
 & 2.244602 & $0.099\pm0.003$ & 2.244602\nl
 & 2.400640 & $0.307\pm0.005$ & 2.399910, 2.400782, 2.401035, 2.401195, 2.401791\nl
 & 2.444659 & $0.087\pm0.005$ & 2.443576, 2.444170, 2.445444\nl
 & 2.528857 & $0.153\pm0.003$ & 2.528504, 2.529211\nl
% & & \nl
{\bf KP 78} & 1.985477 & $0.197\pm0.006$ & 1.985368, 1.985587\nl
 & 2.042732 & $0.155\pm0.008$ & 2.042464$^b$, 2.043000\nl
 & 2.061465 & $0.151\pm0.005$ & 2.061347, 2.061583\nl
 & 2.094603 & $0.912\pm0.004$ & 2.093318, 2.094057, 2.094360, 2.095019, 2.095867\nl
 & 2.097187 & $0.103\pm0.003$ & 2.097187\nl
 & 2.115063 & $0.112\pm0.005$ & 2.115063\nl
 & 2.240122 & $1.620\pm0.012$ & 2.238328, 2.238913, 2.239351, 2.239765, 2.240104, 2.240775,\nl
 &          &                 & 2.241570, 2.242268, 2.242837, 2.243204\nl
 & 2.550918 & $0.131\pm0.004$ & 2.550744, 2.551092\nl
\enddata
\end{deluxetable}
\footnotesize
$^a$apparent broad component, $b$ value uncertain

$^b$highly asymmetric line

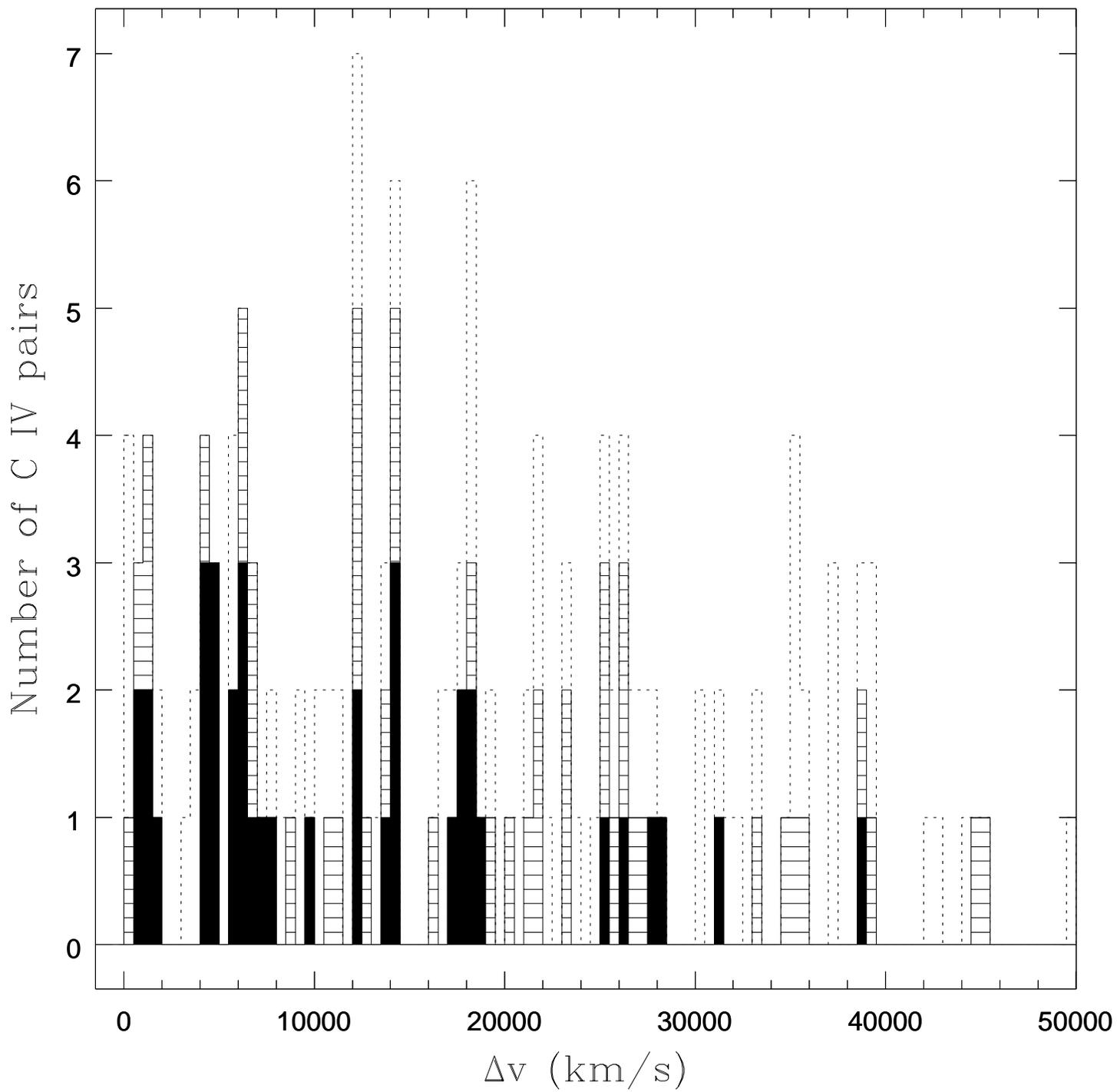
\end{document}